\documentclass[sigconf]{acmart}

\usepackage{booktabs} 

\usepackage{lipsum} 
\usepackage{balance}
\usepackage{paralist}
\usepackage{filecontents}
\usepackage{etoolbox}
\usepackage{mathrsfs}
\usepackage{bm}
\usepackage{setspace}
\usepackage{graphicx}
\usepackage{caption}
\usepackage{subcaption}
\usepackage[ruled]{algorithm2e}
\usepackage{multirow}
\usepackage{amsmath}
\usepackage{hyperref}

\setcopyright{acmcopyright}
\copyrightyear{2018}
\acmYear{2018}
\acmDOI{10.1145/1122445.1122456}

\acmConference[Woodstock '18]{Woodstock '18: ACM Symposium on Neural
	Gaze Detection}{June 03--05, 2018}{Woodstock, NY}
\acmBooktitle{Woodstock '18: ACM Symposium on Neural Gaze Detection,
	June 03--05, 2018, Woodstock, NY}
\acmPrice{15.00}
\acmISBN{978-1-4503-XXXX-X/18/06}

\newcommand{\bb}[1]{\mathbb{#1}}

\begin{document}
\title{A Hierarchical Self-attentive Convolution Network for Review Modeling in Recommendation Systems}

\author{Hansi Zeng}
\affiliation{%
  \institution{School of Computing, University of Utah}
}
\email{hanszenghappy@gmail.com}

\author{Qingyao Ai}
\affiliation{%
  \institution{School of Computing, University of Utah}
}
\email{aiqy@cs.utah.edu}
\begin{abstract}
Using reviews to learn user and item representations is important for recommender system. Current review based methods can be divided into two categories: (1) the Convolution Neural Network (CNN) based models that extract n-gram features from user/item reviews; (2) the Recurrent Neural Network (RNN) based models that learn global contextual representations from reviews for users and items. 
Despite their success, both CNN and RNN based models in previous studies suffer from their own drawbacks. 
While CNN based models are weak in modeling long-dependency relation in text, RNN based models are slow in training and inference due to their incapability with parallel computing. 
To alleviate these problems, we propose a new text encoder module for review modeling in recommendation by combining convolution networks with self-attention networks to model local and global interactions in text together.
As different words, sentences, reviews have different importance for modeling user and item representations, we construct review models hierarchically in sentence-level, review-level, and user/item level by encoding words for sentences, encoding sentences for reviews, and encoding reviews for user and item representations. 
Experiments on Amazon Product Benchmark show that our model can achieve significant better performance comparing to the state-of-the-art review based recommendation models.
\end{abstract}

\maketitle


\section{Introduction}

Review text is considered to be valuable for the learning of effective user and item representations for recommender systems.
Previous studies show that incorporating user reviews into the optimization of recommender systems can significantly improve the performance of rating prediction by alleviating data sparsity problems with user preferences and item properties expressed in review text~\cite{Topic integratedHFT,Topic integratedRMR,DeepCoNN,narre}.
Among different techniques for review modeling in recommendation, deep learning methods, especially convolution nerual networks (CNN) and recurrent neural networks (RNN), are the most popular paradigms in state-of-the-art recommendation models.
For example, DeepCoNN\cite{DeepCoNN} uses two convolution nerual networks to learn user and item representations from reviews; HANN\cite{HANN} uses a sequence of gated recurrent unit(GRU) to model the user and item embeddings from reviews.

Despite their success, both CNN based and RNN based review modeling paradigms have their own weakness. 
While the CNN based models are good at capturing local interactions from text such as n-gram key-phrases, they are weak in learning global interactions from text \cite{rnn and cnn}.
For example, a sentence "\textit{this version is not classic like its predecessor, but its pleasures are still plentiful}" could be classified as having negative sentiment by CNN models because they tend to capture n-grams like ``is not classic'' but not the whole meaning of the sentence. 
While RNN based models are good at learning the semantic meaning of the whole sentence, they suffer from slow training and prediction due to their sequential nature.
Because RNN has to take inputs one by one, it is difficult, if not impossible, to use any parallel computing techniques to speed up their inference in production.

To overcome the shortcomings of CNN and RNN based models, in this paper, we build a new module to encode text information by combining convolution layer with self-attention layer~\cite{attention}. 
Specifically, we use the convolution layer to capture local interaction between words such as key-phrases, and use the self-attention layer to learn global interactions from text to make the whole system friendly to parallel computing.
Also, to capture the hierarchical nature of user/item reviews, we propose to construct multi-layer encoders to capture the review information hierarchically from sentence-level, review-level to user/item-level.
In sentence-level, we encode a sentence from its words by stacking a word-level self-attentive convolution layer and an attention-based aggregation layer to model the contextual representations of words and their informativeness for the sentence. 
In review-level, we learn a review representation from its sentences by stacking the same two layers to capture the local and global interactions between sentences and their corresponding importance for semantic modeling. 
Finally, in user/item-level we learn user and item representations from their reviews using attention networks to select and combine review representations according to their utility for recommendation.
Empirical experiments on four datasets from Amazon Product Benchmark show that our proposed Hierarchical Self-Attentive Convolution Network (HSACN) can extract more effective user and item representations from review data comparing to that state-of-the-art review-based recommendation models.


\section{Related Works}
Using review text to enhance user and item representations for recommender system has been widely studied in recent years \cite{Topic integratedHFT, Topic integratedRMR}. Many works are focus on topic modeling from review text for users and items. For example, the HFT\cite{Topic integratedHFT} uses LDA-like topic modeling to learn user and item parameters from reviews. The likelihood from the topic distribution is used as a regularization for rating prediction by matrxi factorization(MF). The RMR\cite{Topic integratedRMR} uses the same LDA-like model on item reviews but fit the ratings using Guassian mixtures but not MF-like models. 
Recently, with the advance of deep learning, many recommendation models start to combine neural network with review data to learn user and item representations, including DeepCoNN\cite{DeepCoNN}, D-Att\cite{cnndlga}, NARRE\cite{narre}, HUITA\cite{HUITA}, HANN\cite{HANN}. 
Despite their differences, existing work using deep learning models for review modeling can be roughly divided into two categories -- CNN based models and RNN based models. 
For example, DeepCoNN~\cite{DeepCoNN} uses two convolution neural networks to learn user and item representations from reviews; NARRE~\cite{narre} extends CNN with an attention network over review-level to select reviews with more informativeness; HANN~\cite{HANN} encodes reviews using GRU and uses attention networks in word-level and review-level, etc. 
To overcome the existing problems of review modeling with RNN and CNN models, in this paper, we propose a text encoding module by stacking convolution and self-attention layer, where convolution layer captures text local feature and self-attention layer learns text global feature. Also, we use the encoding module hierarchically to flexible pay varying attentions to reviews for user/item representations. 
\section{Proposed Method}
\begin{figure}[t]
    \centering
    \includegraphics[width=0.49\textwidth]{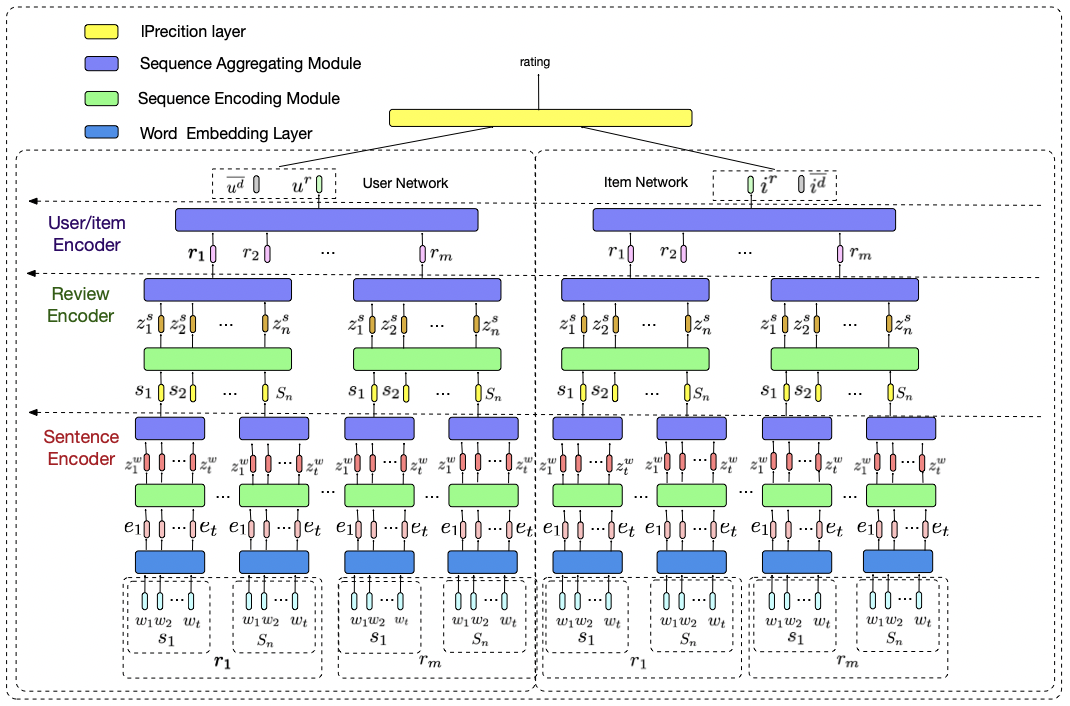}
    \caption{Architecture of the proposed method}
    \label{fig:arch}
\end{figure}

In this section, we introduce our proposed recommendation model based on local convolution and global self-attention, see Figure \ref{fig:arch}. There are three major encoders to encode information: a sentence encoder that learns the embedding of a sentence from its words; a review encoder learning the representation of a review from its sentences; and a user/item encoder which learns user/item representation from its reviews. 
Specifically, we first introduce the sequence encoding and aggregating modules shared by sentence and review encoders, and then the details about other model architectures.

\subsection{Sequence Encoding \& Aggregating Module} \label{sec:module}
\noindent \textbf{\underline{Sequence Encoding Module}} learns a contextual representation for each element in the sequence by considering the local and global relations between them. It is a stack of a convolution layer, a multihead self-attention layer, and a feed-forward layer.  Formally, the input of the module is a sequence of embeddings $\{ \bm{x_1}, \ldots, \bm{x_t} \}$, and the output is $\{ \bm{z_1}, \ldots, \bm{z_t} \}$, where $\bm{x_i} \in \mathbb{R}^d$, $\bm{z_i} \in \mathbb{R}^{d_k}$, $d$ is the dimension of input embeddings, and $d_k$ is the hidden dimension of the module. First, we use the convolution network to learn n-gram features from the input sequence.  Mathematically, for a sequence of embeddings $\{ \bm{x_1}, \ldots, \bm{x_t} \}$, we denote $\bm{c}_i \in \mathbb{R}^{dn}$ as a concatenation of embeddings $\bm{x}_{i-\lfloor \frac{n-1}{2}\rfloor}, \ldots, \bm{x}_{i+\lfloor \frac{n-1}{2}\rfloor}$, where $n$ is the width. Then we can learn the n-gram feature: $\bm{q}_i$, $\bm{k}_i$, $\bm{v}_i$ from  $\bm{x}_{i-\lfloor \frac{n-1}{2}\rfloor}, \ldots, \bm{x}_{i+\lfloor \frac{n-1}{2}\rfloor} $ by using the convolution weights $\{ \bm{W}_Q, \bm{W}_K, \bm{W}_V \} \in \mathbb{R}^{d_k \times nd}$  and biases $\{ \bm{b}_Q, \bm{b}_K, \bm{b}_V\} \in \mathbb{R}^{d_k}$ respectively: 
\begin{align}
    \bm {q}_i &= \text{Relu}(\bm{W}_Q   \bm{c}_i + \bm{b}_Q) \\
\bm{k}_i &= \text{Relu}(\bm{W}_K  \bm{c}_i + \bm{b}_K) \\ 
    \bm{v}_i &= \text{Relu}(\bm{W}_V  \bm{c}_i + \bm{b}_V)
\end{align} \par 
After the extraction of n-gram features, we take them as input to the multihead self-attention network . We use self-attention since it can capture the long-range dependencies of the sequence which is the complementary to local n-gram feature. Formally, we split $(\bm{q}_i, \bm{k}_i, \bm{v}_i )_{i=1}^t$ into H heads, denoted as $( \bm{q}_i^{h}, \bm{k}_i^{h}, \bm{v}_i^{h} )_{i=1}^t$ for each head, where $\{  \bm{q}_i^{h}, \bm{k}_i^{h}, \bm{v}_i^{h} \} \in \mathbb{R}^{d_h}$ and $d_h$ is the subspace dimension for each head. We apply the same operation to each head. Besides we use relative position embeddings \cite{relative attention} $\bm{p}^V$, $\bm{p}^K$ to encode the pairwise distance between input sequences. Formally, each head's output $\bm{z}_i^{h}$ is the weighted sum of transformed inputs with a learnable vector $\bm{p}_{ij}^V \in R^{d_h}$ encoding the distance between element $i$ and element $j$, 
\begin{equation}\mathbf{z}_i^{h} = \displaystyle\sum_{j=1}^t a_{ij}^{h} (\bm{v}^{h}_j + \bm{p}^V_{ij} )\end{equation}
each weighted coefficient is computed using a softmax function,
\begin{equation} a_{ij}^{h} = \frac{\exp{(e_{ij}^{h})}}{\sum_{l=1}^t \exp{(e_{il}^{ h}})}\end{equation}

And $e_{ij}^{h}$ is computed as the inner product between two elements,
\begin{equation}e_{ij}^{h} = \frac{<\bm{q}_i^{h}, \bm{k}_j^{h} + \bm{p}^K_{i,j}>}{\sqrt{d_h}}\end{equation}
where $\bm{p}^K_{i,j} \in \mathbb{R}^{d_h}$ is a learnable parameter. Noted that the $\bm{p}^V_{ij}$ and $\bm{p}^K_{i,j}$ are different parameters since one encodes the positional information between values $(\bm{v}_i^h)_{i=1}^t$, and the other encodes the positional information between keys $(\bm{k}_i^h)_{i=1}^t$, but they can be shared across attention heads.
Finally, we concatenate each head's output $\bm{z}_i^{h}$ to get $\bm{z}_i$ and apply a one layer feed-forward network with weights $\bm{W}_f$, bias $\bm{b}_f$ to learn a more flexible representation,
\begin{equation}
    \bm{z}_i = \text{Relu} (\bm{W}_f \bm{z}_i + \bm{b}_f)
\end{equation}
where $\bm{W}_f \in \bb{R}^{d_k \times d_k}$, $\bm{b}_f \in  \bb{R}^{d_k}$. 

\noindent \textbf{\underline{Sequence Aggregating Module}} takes the output of the sequence encoding module as input, and distills their information to form a single vector using attention network. We then treat the vector as the representation of the sequence. Formally, let $\{ \bm{z_1}, \ldots \bm{z_t} \}$ be the input sequence, then the $i-th$ element importance score is computed as,
\begin{align}
    a_i &= \bm{h}^T \text{Relu} (\bm{W}_p \bm{z}_i + \bm{b}_p) \\
    a_i &= \frac{\exp(a_i)}{\sum_{j=1}^t \exp(a_j)}
\end{align}
where $\bm{W}_p \in \mathbb{{R}}^{d_p \times d_k}$, $\bm{b}_p \in \mathbb{R}^{d_p}$, $\bm{h} \in \mathbb{R}^{d_p}$ are learnable parameters, and in our model, we set $d_p = \frac{d_k}{2}$. 
Then the final output $\bm{l}$ is the weighted sum of contextual representations of each element, 
\begin{equation}
    \bm{l} = \displaystyle\sum_{i=1}^t a_i \bm{z}_i
\end{equation} 
\subsection{Model Architecture}
\textbf{\underline{Sentence Encoder}}. The sentence encoder learns a sentence representation from words, and there are three modules in the encoder.
The first one is a word embedding layer that maps a sequence of words from a sentence to their corresponding dense vectors. Mathematically, assume a sentence $s$ consists of words $\{ w_{1}, \ldots w_{ t}\}$ with size of $t$. The embedding layer transforms the word sequence into dense vectors $\{ \bm{e_1}, \ldots \bm{e_t} \}$, where $\bm{e_i} \in \mathbb{R}^d$. We denote the word embedding matrix as $\mathbf{E} \in \mathbb{R}^{V \times d}$, where $V$ is the word vocabulary size, $d$ is the dense vector dimension. \par 
The second one is a word-level sequence encoding module we defined in section \ref{sec:module}, and we denote it as a function $f_s$. The input of the function is word embeddings $\{ \bm{e_1}, \ldots \bm{e_t} \}$, and the output is their corresponding contextual representations $\{ \bm{z^w_1}, \ldots \bm{z^w_t} \}$, which are computed as,
\begin{equation}
    \{ \bm{z^w_1}, \ldots \bm{z^w_t} \} = f_s (\{\bm{e_1}, \ldots \bm{e_t}\}, d_w)
\end{equation}
where $d_w$ is the hidden dimension of the module, and $\bm{z_i^w} \in \mathbb{R}^{d_w}$. \par 
The third module is a word-level sequence aggregating module defined in section \ref{sec:module}. We denote it as a function $g_s$, and it takes contextual representations $\{ \bm{z^w_1}, \ldots \bm{z^w_t} \}$ as input to get the final sentence embedding, \begin{equation}
    \bm{s} = g_s (\{ \bm{z^w_1}, \ldots \bm{z^w_t} \})
\end{equation}

\noindent \textbf{\underline{Review Encoder}}.
The review encoder learns a review representation from its sentences, which consists of two major modules. The first module is a sentence-level sequence encoding module, and we treat it as a function $f_r$. For each
review, the function takes a sequence of sentence embeddings $\{\bm{s}_1, \ldots \bm{s}_n \}$ as input and get their corresponding contextual representations $\{ \bm{z}^s_1, \ldots \bm{z}^s_n \}$, which are computed as,
\begin{equation}
    \{ \bm{z^s_1}, \ldots \bm{z^s_n} \} = f_r (\{\bm{s_1}, \ldots \bm{s_n}\}, d_s)
\end{equation}
where $d_s$ is the hidden dimension of the the  module, and $\bm{z^s_i}\in \mathbb{R}^{d_s}$.
\par 
The second module of the review encoder is the sentence-level aggregating module, and we denote it as $g_r$. The input of the function is $\{ \bm{z}^s_1, \ldots \bm{z}^s_n \}$, and the output is the review representation $r$, 
\begin{equation}
    \bm{r} = g_r (\{ \bm{z^s_1}, \ldots \bm{z^s_n} \})
\end{equation}
\noindent \textbf{\underline{User/item Encoder}}.
We learn a user representation using review-level aggregating module, denoted as $g_u$, to select their reviews with more informativeness. Formally, we assume a user $u$ writes $m$ reviews
$r_1, \ldots r_m$, then the final user representation can be computed as,
\begin{equation}
    \bm{u}^r = g_u (\{ r_1, \ldots r_m \})
\end{equation}
Also, the representation of item $i$ can be computed in similar way, and we denote it as $\bm{i}^r$. \par 
\noindent \textbf{\underline{Prediction layer}}.
We first project user representation  learned from reviews $u^r$ into the latent space using one layer feed-forward network,
\begin{equation}
    \overline{\bm{u}^r} = \text{Relu}(\bm{W}_u u^r + \bm{b}_u)
\end{equation}
where $\bm{W}_u \in \mathbb{R}^{d_l \times d_s}$, $\bm{b}_u \in \mathbb{R}^{d_l}$. 
Similarly using the equation above, we can get the projected item representation $\overline{\bm{i}^r}$. Also, we introduce another set of embeddings $\overline{\bm{u}^d}$, $\overline{\bm{i}^d}$, with the same dimension as $\overline{\bm{u}^r}$ and $\overline{\bm{i}^r}$, for each user and item, and then the interaction between user and item can be modeled as,
\begin{equation}
    \bm{h} = ( \overline{\bm{u}^d} + \overline{\bm{u}^r} ) \odot ( \overline{\bm{i}^d} + \overline{\bm{i}^r} )
\end{equation}
The final rating prediction can be computed as,
\begin{equation}
    \hat{R}_{uv} = \bm{w}_f^T \bm{h} + b_u + b_v + b_g
\end{equation}
where $\bm{w}_f \in \mathbb{R}^{d_l}$, $b_u, b_v, b_g \in \mathbb{R}$. The $b_u$ represents user bias, $b_v$ is item bias, and $b_g$ is the global bias. \par 
When training model, we minimize the loss between ground truth rating $R_{uv}$ with predicted rating $\hat{R}_{uv}$. We use the mean square error as the loss function, 
\begin{equation}
    \mathcal{L} = \frac{1}{N} \displaystyle\sum_{i=1}^N (R_{uv} - \hat{R}_{uv})^2
\end{equation}
where $N$ is the number of user item pairs in the training dataset.

\section{Experiment}
\noindent \textbf{\underline{Datasets and evluation metrics.}} We conduct our experiment on four different categories of 5-core Amazon product review datasets \cite{amazon dataset}. The domains and statistics of these four categories are shown in the Table \ref{tab:data}. For each dataset, We randomly split user-item pairs into $80\%$ training set,  $10\%$ validation set, and $10\%$ testing set. And we use NLTK\cite{NLTK} to tokenize sentences and words of reviews. Since the length and the number of reviews have a long tail effect, we only keep the number of reviews cover $90\%$ and length of reviews cover $70\%$ users and items for the balance of efficiency and performance. Following\cite{narre}, we adopt Root Mean Square Error(RMSE) as the main metric to evaluate the performance of our model. \par
\begin{table}[t]
    \small  
    \centering
    \caption{Statistical details of the datasets.}
   	\scalebox{0.85}{ 
    \begin{tabular}{|c|c|c|c|c|}
    \hline
     & \textbf{Video Games} & \textbf{Toys and Games} & \textbf{Kindle Store} & \textbf{Office Products} \\ \hline 
     users &  24,303 & 19,412 & 68,223 & 4,905 \\ \hline 
     items  & 10,672 & 11,924 &  61,935  & 2,420 \\ \hline 
     reviews  & 231,780 & 167,597 & 982,619  & 53,258 \\ \hline
    \end{tabular}
     \label{tab:data}
	}
\end{table}

\noindent \textbf{\underline{Baselines.}} To evaluate the performance of our method we compare it with several baseline models. (1)~\textit{NeurMF}: CF based model combines linearity of GMF and  non-linearity of MLPs for modeling user and item latent representations\cite{neurMF}; (2)~\textit{HFT}: topic modeling based model combines the ratings with reviews via LDA\cite{Topic integratedHFT}; (3)~\textit{DeepCoNN}: CNN based model uses two convolution neural network to learn user and item representations \cite{DeepCoNN}; (4)~\textit{D-ATT}: CNN based model extends DeepCoNN by using dual-attention layer over the word-level before convolution\cite{cnndlga}; (5)~\textit{NARRE}: CNN based model modifies the DeepCoNN by using the attention network over review-level to select reviews with more informativeness\cite{narre}. \par

\noindent \textbf{\underline{Parameter settings .}} We use 300-dimensional pretrained word embeddings from Google News\cite{googlenews}, and employ the Adam for optimization with an inital learning rate $0.0002$. We set the hidden dimension of word-level and sentence-level sequence encoding module as $150$, and the latent dimension of the prediction layer as $32$. Also, the convolution kernel size is $3$, and number of head for each self-attention layer is $2$. We apply dropout after the word embedding layer, after each feed forward layer in sequence encoding modules, and before the prediction layer with rate $[0.3, 0.5, 0.5]$. We set the hyper-parameters of baseline models following the settings of their original papers.\par 

\noindent \textbf{\underline{Results and analysis .}} The RMSE results of compared models are shown in Table \ref{tab:result}. 
Several observations can be made from the results. Firstly, models with review information generally outperform models without review information. 
It validates that utilizing rich information from reviews for user and item representations can boost the performance of recommender systems. Secondly, the CNN based models with attention network like NARRE and D-ATT can outperform DeepCoNN consistently. It proves that using attention network to distill important information from reviews can improve the final performance. 
Finally, our model, namely HSACN, shows consistently improvement over all baseline models on all datasets. The relative improvement against NARRE are $1.1\%$ in Video Games, $0.5\%$ in Toys and Games, $2\%$ in Kindle Store and $1.2\%$ in Office Products. The reason might be that 
our model consists of convolution layer with self-attention layer, which can both extract the local and global information of reviews. 
Also, the hierarchical model architecture helps select information more robust and effective. 

\begin{table}[t]
    \small  
    \centering
    \caption{The RMSE scores of the different methods.}
    \scalebox{0.85}{ 
    \begin{tabular}{c|c|c|c|c}
    \hline 
         Dataset &  \textbf{Video Games} & \textbf{Toys and Games} & \textbf{Kindle Store} & \textbf{Office Products}\\ \hline \hline
         NeurMF & 1.0950 & 0.9359 & 0.8246 & 0.8809 \\ \hline
         HFT & 1.0834 & 0.9075 & 0.7980 & 0.8452 \\ \hline 
         DeepCoNN & 1.0737 & 0.8950 & 0.7938 &  0.8412 \\ \hline
         D-ATT & 1.0667 & 0.8914 & 0.7905 & 0.8335 \\ \hline 
         NARRE & 1.0630 &  0.8932 & 0.7881 & 0.8331 \\ \hline \hline
        HSACN & \textbf{1.0511} & \textbf{0.8871} & \textbf{0.7721} & \textbf{0.8230}  \\ \hline
    \end{tabular}
	}
    \label{tab:result}
\end{table}

\begin{figure}[h]
    \centering
        \includegraphics[width=0.45\textwidth]{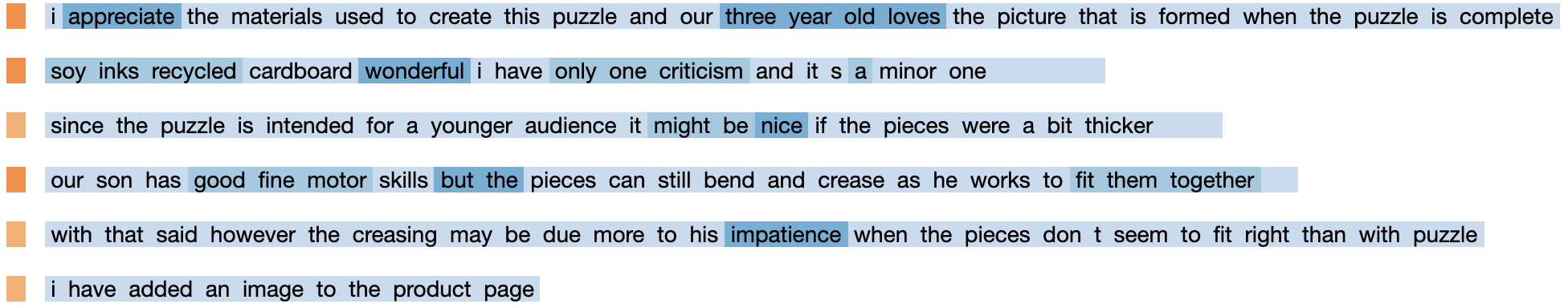}
    \caption{Randomly select a (user,item) pair on the Toys and Games dataset, then visualize a most important review chosen by review-level sequence aggregating module. Orange boxes to the left represent the sentence important score, and blue boxes on each individual word represent the word important score. Darker color means higher important score.}
    \label{fig:visualize}
\end{figure}

\noindent \textbf{\underline{Case study}}. We conduct a case study to check if our model can select informative words and sentences. The visualization of the attention weights in word-level and sentence-level aggregating module are shown in Figure \ref{fig:visualize}. The ground-truth rating is $4.0$. We can see that the word-level aggregating module can not only effectively select the positive words like "appreciate", "loves", "wonderful", "nice", but also sentiment negation words like "criticism", "but". And with the help of self-attention layer of our model that can capture global interactions of text, the predicted rating is $4.23$ which is roughly a positive rating but considers the effect of negative sentiment words. Also the important sentences can be effectively selected by the sentence-level aggregation module. For example, the sentence "I appreciate the materials used to the puzzle and our three year old loves the picture that is formed when the puzzle is complete" is assigned higher score, since it can more effectively represent the user preferences and item properties. In contrast, the sentence 
"I have added an image to the product page" has lower attention score since it contains limit information of users and items.

\section{Conclusion}
In this paper, we propose a hierarchical self-attentive convolution network (HSACN) to learn user and item representations from reviews in recommendation systems. We propose to use a text encoder combining convolution networks with self-attention networks to learn local and global interactions in text together, and construct our model hierarchically in sentence-level, review-level and user/item level. Experiments and case studies show that our model can achieve better performance and extract more effective user and item representations from reviews compared with other state-of-art review based recommendation models. 
For future work, we want to conduct more analysis on the performance and efficiency of our model in other recommendation scenarios such as ranking and diversification.



\end{document}